\documentclass[prl,aps,preprint,amsmath,superscriptaddress]{revtex4-1}

\usepackage{graphicx}
\usepackage{dcolumn}
\usepackage{bm}
\usepackage{latexsym,amsmath,amssymb}
\usepackage{textcomp}
\usepackage{gensymb}
\usepackage{threeparttable}
\usepackage{booktabs}

\begin{document}

\title{Diffusive Pseudo-Conformal Mapping: Anisotropy-Free Transformation Thermal Media with Perfect Interface Matching}
	
\author{Gaole Dai}\email{gldai@ntu.edu.cn}
\affiliation{School of Sciences, Nantong University, Nantong 226019, China}
	
\author{Fubao Yang}
\affiliation{Department of Physics, State Key Laboratory of Surface Physics, and Key Laboratory of Micro and Nano Photonic Structures (MOE), Fudan University, Shanghai 200433, China}
	
\author{Jun Wang} 
\affiliation{School of Physics, East China University of Science and Technology, Shanghai 200237, China}
\affiliation{Wenzhou Institute, University of Chinese Academy of Sciences, Wenzhou 325001, China}
	
\author{Liujun Xu}\email{ljxu@gscaep.ac.cn}
\affiliation{Graduate School of China Academy of Engineering Physics, Beijing 100193, China}

\author{Jiping Huang}\email{jphuang@fudan.edu.cn}
\affiliation{Department of Physics, State Key Laboratory of Surface Physics, and Key Laboratory of Micro and Nano Photonic Structures (MOE), Fudan University, Shanghai 200433, China}
	
\date{\today}
	
\begin{abstract}
Transformation media provide a fundamental paradigm for field regulation, but their tricky anisotropy challenges fabrication. Though optical conformal mapping has been utilized to eliminate anisotropy, two key factors still hinder its development in thermotics, i.e., the distinct diffusion nature and inevitable interface mismatching. Here, we put forth the concept of diffusive pseudo-conformal mapping, overcoming the inherent difference between diffusion and waves and achieving perfect interface matching. The proposed mapping directly leads to heat guiding and expanding functions with anisotropy-free transformation thermal media, whose feasibility is confirmed by experiments or simulations. Besides diverse applications, we provide a unified perspective for two distinct types of prevailing bilayer cloaks by uncovering their profound ties with pseudo-conformal mapping. These results greatly simplify the preparation of transformation thermotics and have implications for regulating other diffusion and wave phenomena.
\end{abstract}

\maketitle

Heat control is essential for every aspect of human life, such as energy utilization, chip cooling, and infrared detection. The past decade has witnessed the development of transformation theory~\cite{Science06-02,Science06-01} in heat conduction, a diffusion process that intrinsically differs from wave dynamics~\cite{NRM2021,PR2021}. Transformation thermotics indicates that anisotropic and inhomogeneous thermal parameters in physical space can mimic heat transfer in curved space, ensured by the form-invariance of heat equations under coordinate transformations. However, anisotropy is a considerable restriction on practical realization because natural materials usually exhibit isotropic thermal properties. Though alternative schemes like diffusive scattering cancellation were proposed to avoid anisotropy~\cite{PRL2014B,PRL2014A,NM2019,epl2020}, they generally apply to specific scenarios such as thermal invisibility. Therefore, removing the intrinsic anisotropy of transformation thermal media is still a big challenge.

Conformal transformation optics~\cite{Science06-02,NP2015} can eliminate anisotropy with two-dimensional (2D) optical conformal mapping that locally preserves the angles and orientations of curves~\cite{castein}. Diverse wave phenomena have been realized with anisotropy-free transformation refractive index~\cite{Science06-03,nano,prx15,nc2022,zsn,np2016,sunfei,hyprap19,nc21}. However, two critical problems challenge the application of optical conformal mapping in thermotics. On the one hand, diffusion and waves are fundamentally different in governing equations (i.e., the Laplace equation vs. the Helmholtz equation) and key parameters (i.e., thermal conductivity vs. refractive index). On the other hand, matching the interface heat flux between the functional device and background is still tricky due to the lack of an intuitive impedance matching criteria, which is essential for accurate and robust heat manipulation. Thus, directly utilizing optical conformal mapping for precise diffusion control is impracticable.

Here, we propose the concept of diffusive pseudo-conformal mapping with angle preservation for certain families of curves representing thermal fields. Fortunately, ``pseudo'' does not affect the crucial advantage of ``conformal'' in removing anisotropy and contributes to perfect interface matching. The proposed mapping yields precise heat manipulation with anisotropy-free transformation thermal media, and we take heat guiding and expanding as two typical examples, with experimental or simulated confirmation. Moreover, we reveal the geometric origin of bilayer cloaks previously designed by scattering cancellation from the perspective of pseudo-conformal mapping. These results feature scalability in handling complex heat transfer~\cite{PRL2015,PRAP2022yang,PRE2018} and provide a unified geometric perspective towards various thermal functions with isotropic materials.

For clarity, we first establish diffusive conformal mapping and illustrate its restrictions. A 2D conformal mapping $f: z_0\mapsto z$ ($z_0=x_0+i y_0$ and $z=f(z_0)=x+i y$) is holomorphic, satisfying the Cauchy–Riemann equations~\cite{castein},
\begin{equation}\label{cr}
	\frac {\partial f}{\partial {\overline {z_0}}}={\frac {1}{2}}\left({\frac {\partial f}{\partial x_0}}+i{\frac {\partial f}{\partial y_0}}\right)=0,
\end{equation}
where $\overline {z_0}$ is the complex conjugate of $z_0$. 
We consider form-invariant heat conduction equations in physical space (position denoted by $z$) and virtual space (position denoted by $z_0$),
\begin{subequations}
	\begin{align}
    \begin{split} 
	\nabla\cdot\left(\boldsymbol\kappa\nabla T(z)\right)-Q=0,	
    \end{split}\\
    \begin{split}
	\nabla_0\cdot\left(\boldsymbol\kappa_0\nabla_0 T_0(z_0)\right)-Q_0=0,
	\end{split}
	\end{align}
\end{subequations}
where $T$, $\boldsymbol \kappa$, and $Q$ are the temperature, thermal conductivity (rank-2 tensor), and internal source power in physical space, respectively. Their counterparts with a subscript ``0'' denote corresponding parameters in virtual space.
The transformation rules to ensure $T(z)=T_0(f^{-1}(z_0))$ are~\cite{61,153}
\begin{subequations}
	\begin{align}
		\begin{split}\label{trans-fuliye}
		\boldsymbol \kappa(z)=\text{J}_f\boldsymbol\kappa_0(z_0)\text{J}_f^{\operatorname {T}}/\det\text{J}_f,	
		\end{split}\\
		\begin{split}\label{trans-q}
		Q(z)=Q_0(z_0)/\det\text{J}_f,
		\end{split}
	\end{align}
\end{subequations}
where $\text{J}_f$ is the Jacobian matrix of $f$, and $\text{J}_f^{\operatorname {T}}$ is the transpose of $\text{J}_f$. Due to the holomorphicity of $f$, $\text{J}_f \text{J}_f^{\operatorname {T}}/\det\text{J}_f$ is an identity matrix~\cite{NP2015}, so Eq.~(\ref{trans-fuliye}) can be simplified as
$\boldsymbol \kappa(z)=\boldsymbol\kappa_0(z_0(z))$.
With the familiar paradigm of transformation theory, virtual space is isotropic and homogeneous, so $\kappa$ and $\kappa_0$ are identical constant scalars. This result is consistent with the conventional technique using conformal mapping to solve heat equations in irregular geometric domains~\cite{handbook2}.
Different from the engineered gradient index in conformal transformation optics, such a frozen degree of freedom (i.e., $\kappa=\kappa_0$) restricts the function design for thermal manipulation.
Besides, transformation media are often in contact with the background that undergoes trivial mapping without changing material properties. However, conformal mapping usually cannot ensure the continuity of boundary conditions due to the strict constraint of Eq.~(\ref{cr}), i.e., conformality for all curves, as shown in the examples below. In fact, in most cases, we do not even have the option to consider interface matching due to the uniqueness of conformal mapping between two given domains. Therefore, we must develop diffusive pseudo-conformal mapping further  and design thermal guiding and expanding functions.

\begin{figure} 
	\centering
	\includegraphics[width=0.7\linewidth]{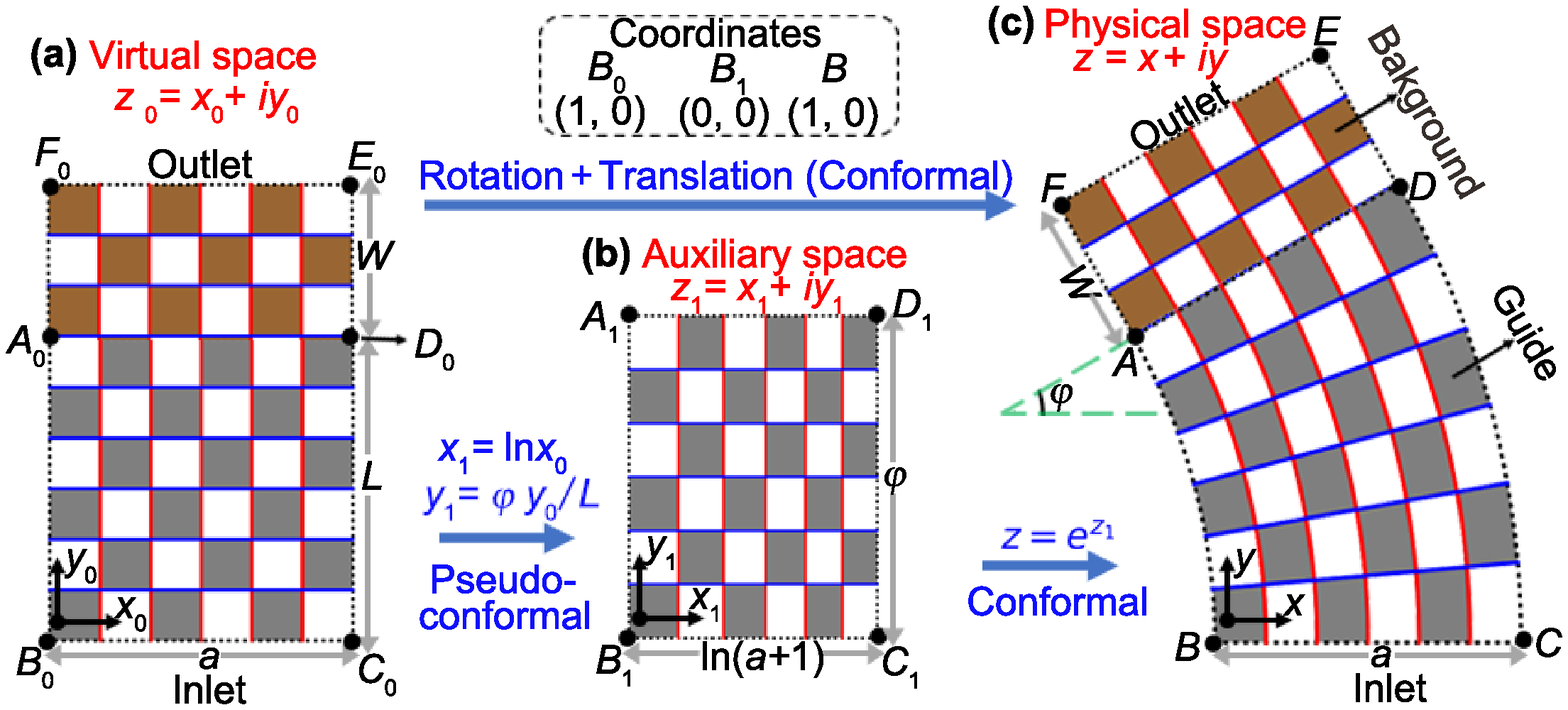}
	\caption{Transformation for a heat flux guide (gray and white meshes) including the background (brown and white meshes). (a), (b) and (c) are the virtual space, the auxiliary space (only showing the transformation of the guide) and the physical space, respectively.}\label{f2}
\end{figure}

Thermal guiding can bend the heat flux by an arbitrary angle. Its virtual space and physical space are shown in Figs.~\ref{f2}(a) and \ref{f2}(c), respectively. In the virtual space, we apply a thermal bias  along the $y_0$-axis. The inlet (hot source) is put on the bottom horizontal boundary $B_0C_0$ and the outlet (cold source) is on the top boundary $F_0E_0$. The vertical boundaries $F_0A_0B_0$ and $E_0D_0C_0$ are thermally insulated.
In the physical space, we expect the flux is rotated by angle $\varphi$ when flowing out of the guide without changing its magnitude. In Fig.~\ref{f2}(a), we plot the grid lines of the Cartesian coordinates.
They are just the heat flux streamlines (constant-$x_0$ curves) and the isotherms (constant-$y_0$ curves).
The upper rectangle (with a height $W$) undergoes a composition of rotation and translation to the background in Fig.~\ref{f2}(c) without changing its thermal conductivity.  
The lower rectangle (with a height $L$) is transformed into the partial annulus (the guide) in Fig.~\ref{f2}(c). Notably, an identity transformation should happen between the inlets $B_0C_0$ and $BC$. Also, the mappings for the background and the guide should have the same effect at their interface $AD$ (on the constant-azimuth line equal to $\varphi$ in Fig.~\ref{f2}(c), which is mapped from $A_0D_0$ in Fig.~\ref{f2}(a)). 
The other boundaries, $FAB$ and $EDC$, are still insulated in the physical space. Although there exists a conformal mapping  between the guide and its preimage according to Riemann mapping theorem~\cite{castein}, it generally cannot transform $B_0C_0$ (or $A_0D_0$) into $BC$ (or $AD$), let alone achieve the interface effect as we want (see our discussion based on the theory of quasiconformal mapping in Supplemental Material, Note I~\cite{sup}).

To construct the guide, a two-step approach is implemented. A non-conformal mapping is first used [Figs.~\ref{f2}(a) and \ref{f2}(b)]: $x_1=\ln x_0$ and $y_1=\varphi y_0/L$, from the virtual space to an intermediate named the auxiliary space (position denoted by $z_1=x_1+iy_1$).
The second step to the physical space [Figs.~\ref{f2}(b) and \ref{f2}(c)] is conformal: $z=\exp(z_1)$, which can lead to the waveguide in optics~\cite{NP2015}. The composition transformation is also non-conformal and the thermal conductivities in the auxiliary space (denoted by $\boldsymbol \kappa_1$) and the physical space (satisfying $\boldsymbol \kappa(z)=\boldsymbol \kappa_1(z_1)$) should be diagonally anisotropic, writing $\boldsymbol \kappa=\kappa_0 \text{diag}\left(\frac{\partial x_1}{\partial x_0}/\frac{\partial y_1}{\partial y_0},\frac{\partial y_1}{\partial y_0}/\frac{\partial x_1}{\partial x_0}\right)$ in the Cartesian coordinate system. 
However, if $\boldsymbol \kappa_0$ itself is diagonally anisotropic, e.g., $\boldsymbol \kappa_0\sim \text{diag}\left(\frac{\partial y_1}{\partial y_0}/\frac{\partial x_1}{\partial x_0},\frac{\partial x_1}{\partial x_0}/\frac{\partial y_1}{\partial y_0}\right)$, anisotropy can be eliminated in the other two spaces. In addition, since only $\kappa^{y_0y_0}_0$ contributes to heat flux, we can take $\kappa_0=\kappa^{y_0y_0}_0$ without changing the results in any space. In this way, thermal conductivities in all spaces are isotropic and we have
\begin{equation}\label{guide}
	\kappa(z)=	\kappa_1(z_1(z))=\kappa_0\dfrac{\varphi}{L}\sqrt{x^2+y^2}.
\end{equation}
We also plot the grid lines in Figs.~\ref{f2}(b) and \ref{f2}(c) mapped from their counterparts in Fig.~\ref{f2}(a). The three sets of grid lines are all orthogonal since they happen to be streamlines and isotherms in their spaces.  
Intuitively, we call the mapping in the first step ``pseudo-conformal'' for maintaining orthogonality of certain curves. By doing a pseudo-conformal mapping first and then a conformal one, the composition is still pseudo-conformal. The streamlines in Figs.~\ref{f2}(a) and \ref{f2}(c), which are both evenly distributed, so the heat flux magnitude is not changed. Also, the flux in the guide does turn an angle $\varphi$ as we expected.

\begin{figure}
	\centering
	\includegraphics[width=0.50\linewidth]{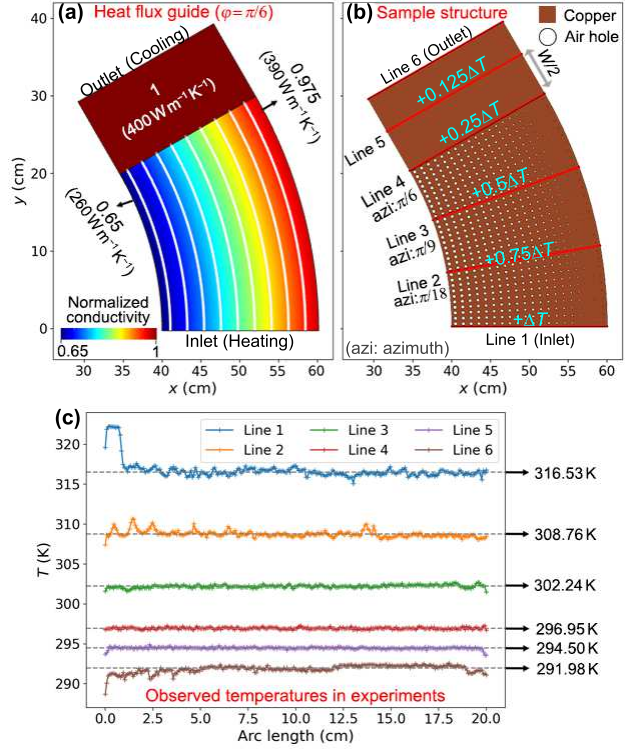}
	\caption{(a) Normalized thermal conductivity ($\kappa/\kappa_0$) profile of the guide. The white curves are isolines. (b) Sample structure made of copper and air holes for experimental setup. (c) Measured temperatures showing the data on the red lines plotted in (b). The arc length means the distance from left endpoint.}\label{f22}
\end{figure}

To confirm our design, wo perform an experiment with the following parameters: $a=0.5$, $L=3W=\varphi/0.65$, $\varphi=\pi/6$ and $\kappa_0=400$~W m$^{-1}$ ~K$^{-1}$ (e.g., copper). Here, the unit length is 40~cm. The thermal conductivity profile based on Eq.~(\ref{guide}) is shown in Fig.~\ref{f22}(a) and the background has the same value as $\kappa_0$. The inhomogeneous $\kappa$ is approximated by a composite of copper and air holes [Fig.~\ref{f22}(b)]. We plot six straight lines, including the inlet (Line 1) and outlet (Line 6) in Fig.~\ref{f22}(b). Line 4 is the interface of the guide and background. Lines 2 and 3 divide the guide into thirds, and Line 5 is in the middle of the background. Ideally, the temperature difference relative to the outlet on each line (isotherm) is marked in Fig.~\ref{f22}(b). The total thermal bias is $\Delta T$, which is experimentally realized by water baths heating or cooling the sample. Fig.~\ref{f22}(c) shows the observed temperatures. Each of the six lines is roughly isothermal, and we plot horizontal lines corresponding to their mean value (excluding edge extremes). Notably, the temperature differences between Lines 4-6 are almost equal. We can conclude that the sample can bend the flux and keep its homogeneity. More details about the experimental setup and data are presented in Supplemental Material, Note II~\cite{sup}.

\begin{figure}
	\centering
    \includegraphics[width=0.7\linewidth]{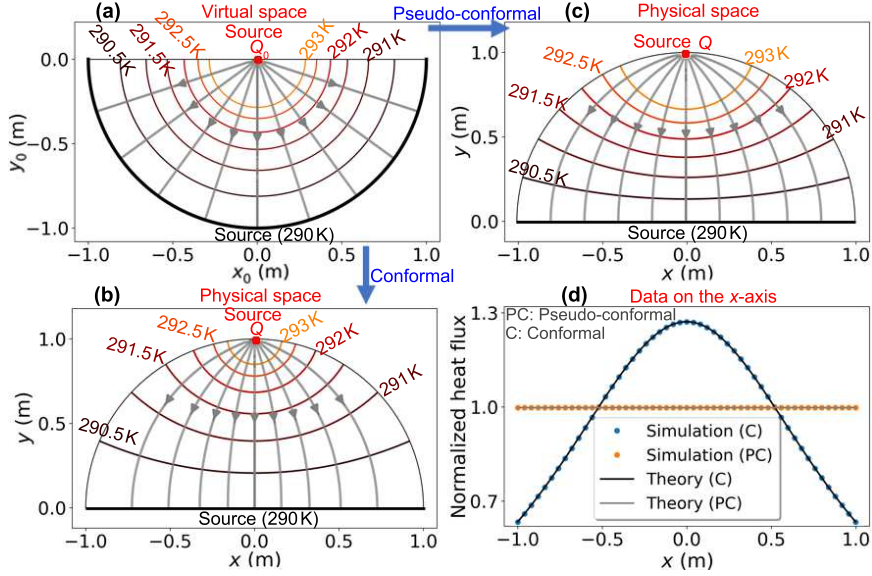}
    \caption{Thermal expander. (a) The virtual space with the point source at the origin. Concentric semicircles are isotherms. Gray curves with arrows represent streamlines, which are also constant-azimuth lines ranging from $1.1\pi$ to $1.9\pi$ ($0.1\pi$ interval from left to right).  (b) The physical space for a conformal mapping. (c) The physical space for a pseudo-conformal mapping. The thermal fields in (a)--(c) are illustrated based on finite element numerical results. (d) The magnitude of the normalized heat flux on the $x$-axis for the two physical spaces. The solid lines are theoretical results while the scatter charts with markers are numerical results. Here, we take $\kappa_0=400$~W~m$^{-1}$ K$^{-1}$ and $Q_0=3000$~W~m$^{-3}$.}\label{fe}
\end{figure}

Next, we design a thermal expander that can convert the heat flux emitted by the point source into parallel flows. In the virtual space [Fig.~\ref{fe}(a)], we consider the lower half-plane. The isotherms form concentric semi-circles with the point source at the center, and the heat flux magnitude  varies with azimuth. To determine the temperature distribution, its value on a certain isotherm should be given, e.g., 290~K on $\{z_0:|z_0|=1, y\leqslant 0 \}$ via an external source. The $x_0$-axis is thermally insulated  except for the point source.
By doing a conformal mapping $z=i\frac{i+z_0}{i-z_0}$ used in the optical counterpart~\cite{NP2015}, i.e., a Möbius transformation, the unit lower half-disk  becomes the unit upper half-disk in the physical space [Fig.~\ref{fe}(b)]. The point source is now at $z=i$ and its power becomes $Q=Q_0/2$ to ensure $T(z)=T_0(z_0(z))$.
The diameter on the $x$-axis $\{z:|x|\leqslant1, y= 0 \}$  is an isotherm (and also the external source) mapped from $\{z_0:|z_0|=1, y\leqslant 0 \}$ and the heat flux on it  is parallel (along the negative $y$-axis).  However, flux magnitude has a varying value depending on which azimuth  it is mapped from.  For practical applications, we want a homogenized flux so it can keep parallel in a rectangular extension attached to $\{z:|x|\leqslant1, y= 0 \}$ when the external source is moved to the bottom of the extension.   
This homogenization can be realized by $z=i\frac{i+z_1(z_0)}{i-z_1(z_0)}$. Here, $\text{Arg}[z_1]=2\arctan \left(\frac{\text{Arg}[z_0]-2\pi}{\text{Arg}[z_0]-\pi}\right)+2\pi$ is pseudo-conformal from the virtual space to the auxiliary space, and Arg denotes the argument (see Supplemental Material, Note III for how to construct this mapping~\cite{sup}).  This approach leads to the same boundary conditions of external source and insulation as the conformal one. Now, the thermal conductivity 
in the expander is
\begin{equation}\label{expander0}
	\kappa(x,y)=\kappa_1\left(1+\sqrt{\frac{\left(x^2+y^2-1\right)^2}{{x^4+2 x^2 \left(y^2+1\right)+\left(y^2-1\right)^2}}}\right)^{-1}.
\end{equation}
 Since $\kappa$ is still isotropic, the heat flux on the $x$-axis can keep parallel.
In Fig.~\ref{fe}(c), we plot the thermal fields in the new physical space. We can see that the intersections of streamlines with the $x$-axis are now uniformly distributed, which is different from the case in Fig.~\ref{fe}(b). This is an intuitive representation of homogenization.
We further show the heat flux on the $x$-axis produced by the two mappings in Fig.~\ref{fe}(d). The theoretical results agree with the finite-element numerical ones from COMSOL Multiphysics (https://www.comsol.com/). In Supplemental Material, Note IV~\cite{sup}, we confirm the performance of our design when considering an extension. Eq.(\ref{expander0}) can also be written in a compact form using transposed bipolar coordinates (See Supplemental Material, Note V~\cite{sup}).

\begin{figure}
	\centering
	\includegraphics[width=0.7\linewidth]{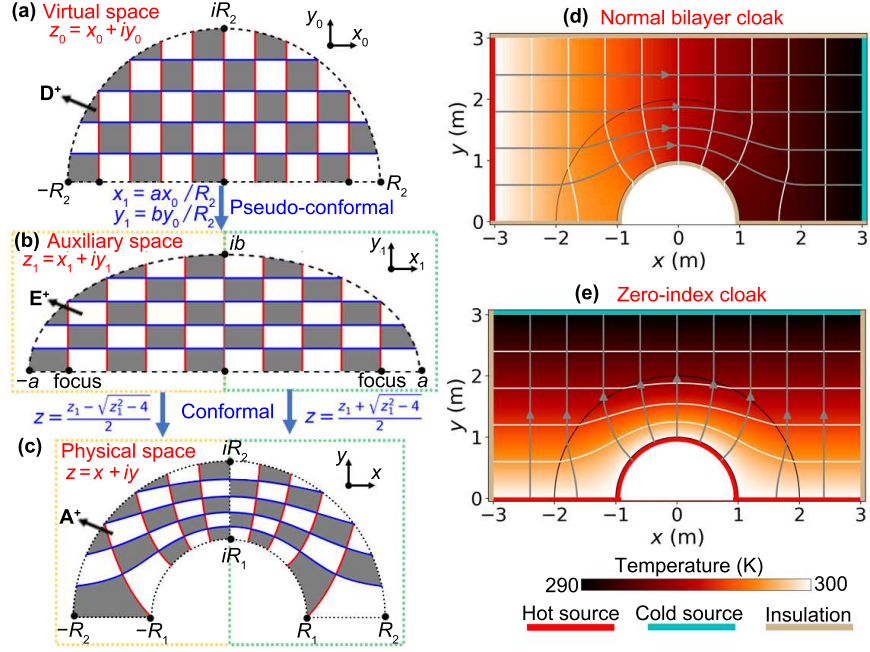}
	\caption{Carpet cloaks. (a)--(c) show the geometric transformation to construct such a cloak. (a), (b), and (c) are the virtual space, auxiliary space and the physical space, respectively. (d) is the computed temperature profile of a normal bilayer cloak with an insulating  inner layer. We place the hot source (300~K) and the cold source (200~K) on the left and right boundaries, respectively, generating a thermal bias along the $x$-axis. (e) is the computed temperature profile of a zero-index cloak with an constant-temperature inner layer (realized by an external source). We illustrate white curves for isotherms and gray curves with arrows for streamlines. Here, we take $R_2=2R_1=2$~m and $\kappa_0=400$~W~m$^{-1}$ K$^{-1}$. The entire simulation domain is limited to a 6~m~$\times$~3~m rectangle.}\label{f1}
\end{figure}

Besides functional design in different scenarios, our approach can also reveal the underlying ties between material parameters and coordinate transformations for some previous works based on alternative schemes of transformation theory.
Here, for example, we provide a geometric insight of
bilayer cloaks~\cite{PRL2014A,PRL2014B,NM2019,epl2020}. Transformation-based annular cloak (shell cloak) depends on a non-homeomorphic mapping to generate its inner boundary, e.g., transforming a point or a line into a closed curve~\cite{Science06-01,Science06-02}.
Due to the geometric symmetry, we consider the upper half of an annular cloak, i.e., a carpet cloak.
The virtual and physical spaces in Fig.~\ref{f1} show a transformation from the upper half-disk $\mathbf {D}^{+}$ with a radius $R_2$ to the upper half-annulus $\mathbf {A}^{+} =\{z:R_1\leqslant|z|\leqslant R_2, y\geqslant 0 \}$ [Fig.~\ref{f1}(c)]. 
Here $\mathbf {A}^{+}$ is actually the outer layer of a bilayer cloak.
For simplicity, we take $R_1=1$ (in meters).  The area outside $\mathbf {D}^{+}$ or  $\mathbf {A}^{+}$ is the background. A cloak means the area enclosed by $\mathbf {A}^{+}$ is expected to have no disturbance to the background.
We still use a two-step pseudo-conformal mapping. First, $\mathbf {D}^{+}$ is compressed/expanded into an half-ellipse $\mathbf {E}^{+} =\{z_1:0\leqslant x_1^2/a^2+y_1^2/b^2  \leqslant 1, y_1\geqslant 0 \}$ [Fig.~\ref{f1}(b)] by  $x_1=\frac{R_2}{a}x_0$ and $y_1=\frac{R_2}{b}y_0$. 
Here we take $a^2=R_2^2+R_1^2$ and $b^2=R_2^2-R_1^2$ so the foci of the half-ellipse are $(\pm 2,0)$. 
This non-conformal  mapping is angle-preserving for the Descartes grid lines in Fig.~\ref{f1}(a).
Second, conformally map $\mathbf {E}^{+}$ to $\mathbf {A}^{+}$:
\begin{equation}\label{iz}
		\left\{\begin{array}{l}z=\dfrac{z_1-\sqrt{z_1^2-4}}{2}, \quad {\text{if }} x_1<0\\ z=\dfrac{z_1+\sqrt{z_1^2-4}}{2}, \quad {\text{if }} x_1\geqslant 0.\end{array}\right.
\end{equation}
This mapping is  one branch of the inverse  Zhukovsky transform~\cite{ca} (See Supplemental Material, Note VI for detailed explanation~\cite{sup}). 
In particular, the upper boundary of $\mathbf{A}^{+}$ (i.e., $\{z:|z|=R_2, y\geqslant 0 \}$) finally undergoes an identity transformation as well as the background.
If the thermal bias is applied along the $x_0$-axis, the grid lines in Fig.~\ref{f1}(a)  represent horizontal streamlines and vertical isotherms and only $\kappa^{x_0x_0}_0$ contributes to heat flux.
Taking $\kappa_0=\kappa^{x_0x_0}_0$, we find the thermal conductivity in $\mathbf{A}^{+}$ is
\begin{equation}\label{aa}
		\kappa=\dfrac{R_2^2+R_1^2}{R_2^2-R_1^2}\kappa_0,
\end{equation}
which is just the cloaking condition for the outer layer of a bilayer cloak~\cite{PRL2014B}. In addition,  the lower boundary of $\mathbf {A}^{+}$ including $\{z:|z|=R_1, y\geqslant 0 \}$ and $\{z:R_1\leqslant|x|\leqslant R_2, y=0 \}$ (denoted by $\Gamma$ as a whole) are mapped from a streamline in the virtual space. The heat flux should have no normal component on $\Gamma$, which can be ensured by the insulation condition. This corresponds to the cloaking condition for the inner layer of a bilayer cloak, i.e., zero-value thermal conductivity and arbitrary shape as along as it covers $\{z:|z|=R_1, y\geqslant 0 \}$. 

If the thermal bias is instead along the $y_0$-axis, our mapping is still angle-preserving for the vertical streamlines and horizontal isotherms  in Fig.~\ref{f1}(a). Take $\kappa_0=\kappa^{y_0y_0}_0$ and we have
\begin{equation}
	\kappa=\dfrac{R_2^2-R_1^2}{R_2^2+R_1^2}\kappa_0.
\end{equation}
The lower boundary $\Gamma$ is mapped from an isotherm, leading to another bilayer cloak made of zero-index materials~\cite{NM2019}. We call the  previous cloak ``normal bilayer'' to distinguish them. The term ``zero-index'' refers to the constant-temperature condition on the inner layer that can be replaced by an effectively infinite thermal conductivity~\cite{NM2019,epl2020}. Figs.~\ref{f1}(d) and \ref{f1}(e) numerically confirm the performance of the two cloaks. In addition to the invisibility effect, we can see that the patterns of isotherms and streamlines in the two cloaks have a duality relationship by swapping the family of curves of streamlines and isotherms.
Further, this  mapping can  be performed on the entire complex plane to build  the shell cloak.  Our method can also be used to design invisibility devices with other geometries, such as confocally elliptical cloaks (See Supplemental Material, Note II for detailed discussions~\cite{sup}).

In summary, we propose the concept of diffusive pseudo-conformal mapping to simultaneously achieve precise heat flux regulation and maintain the material isotropy in transformation thermotics.
By preserving the angles of certain families of curves (e.g., isotherms and streamlines), our approach can circumvent anisotropy and perfectly match the interface heat flux between the transformation media and background.
We demonstrate our theory by designing feasible and robust thermal devices that can bend or parallelize heat flux at our will. Also, we revisit scattering-cancellation-based bilayer cloaks from a unified perspective of pseudo-conformal mapping.
In addition to reobtaining the parameters without inversely solving heat equations, the intrinsic geometric relationship between different types of bilayer cloaks is also revealed.
The idea of diffusive pseudo-conformal mapping can be further developed for controlling transient heat conduction~\cite{153,NSR2023}, multithermotics~\cite{PRAP2019,zaitou}, and other diffusion physics~\cite{science2014,prl20143,Science2012,ciprap}. The consideration of perfect interface matching also benefits the design of transformation wave media, such as avoiding impedance mismatch that plagues conformal invisibility devices~\cite{hyprb}.

We acknowledge financial support from the National Natural Science Foundation of China (Grants No. 11725521, No. 12035004, No. 12147169, and No. 12205101) and the Science and Technology Commission of Shanghai Municipality (Grant No. 20JC1414700).

\clearpage
\newpage

\clearpage
\newpage

\clearpage
\newpage

\clearpage
\newpage

\end{document}